\documentclass[10pt,conference]{IEEEtran}
\makeatletter
\def\ps@headings{%
\def\@oddhead{\mbox{}\scriptsize\rightmark \hfil \thepage}%
\def\@evenhead{\scriptsize\thepage \hfil \leftmark\mbox{}}%
\def\@oddfoot{}%
\def\@evenfoot{}}
\makeatother
\IEEEoverridecommandlockouts
\usepackage{cite}
\usepackage{amsmath,amssymb,amsfonts}
\usepackage{algorithmic}
\usepackage{graphicx}
\usepackage{textcomp}
\usepackage{comment}
\usepackage[T1]{fontenc}
\usepackage{longtable}
\usepackage{lscape}
\usepackage{caption}
\usepackage{float}
\usepackage[caption=false,font=footnotesize]{subfig}
\usepackage{multirow}
\usepackage{graphicx}
\usepackage[table,xcdraw]{xcolor}
\usepackage[normalem]{ulem}

\def\BibTeX{{\rm B\kern-.05em{\sc i\kern-.025em b}\kern-.08em
    T\kern-.1667em\lower.7ex\hbox{E}\kern-.125emX}}

\definecolor{darkgreen}{rgb}{0.29, 0.33, 0.13}
\definecolor{std_red}{rgb}{1, 0, 0}
\definecolor{std_green}{rgb}{0, 1, 0}

\usepackage[textwidth=17mm]{todonotes}
\newcommand{\customtodo}[4]{
        \todo[color=#2,inline,size=\small]{
                \ifx&#3&
                        \textbf{#1} #4
                \else
                        \textbf{#1$\Rightarrow$#3} #4
                \fi
        }
}

\newcommand{\system}{XRP-NDN Overlay}
\begin{document}
\bstctlcite{IEEEexample:BSTcontrol}

\title{To Squelch or not to Squelch: Enabling Improved Message Dissemination on the XRP Ledger
%\thanks{Acknowledgment: we thankfully acknowledge the support from the RIPPLE University Blockchain Research Initiative for our research}
}

%% OLD AUTHORS BLOCK %%%
\begin{comment}
\author{%
\IEEEauthorblockN{Lucian Trestioreanu}
\IEEEauthorblockA{\textit{Interdisciplinary Centre for}\\\textit{Security, Reliability and Trust}\\\textit{University of Luxembourg}\\
Luxembourg, Luxembourg \\
lucian.trestioreanu@uni.lu}
\and \IEEEauthorblockN{Wazen Shbair}
\IEEEauthorblockA{\textit{Luxembourg Company} \\
\textit{Luxembourg}\\ Luxembourg \\ wazen.shbair@gmail.com}
\and \IEEEauthorblockN{Flaviene Scheidt}
\IEEEauthorblockA{\textit{Interdisciplinary Centre for}\\\textit{Security, Reliability and Trust}\\\textit{University of Luxembourg}\\
Luxembourg, Luxembourg \\
flaviene.scheidt@uni.lu}
%\and \IEEEauthorblockN{Aanchal Malhotra}
%\IEEEauthorblockA{\textit{Xpring} \\
%\textit{Northeastern University}\\ Boston, USA \\ %amalhotra@ripple.com}
\and  \IEEEauthorblockN{Radu State}
\IEEEauthorblockA{\textit{Interdisciplinary Centre for}\\\textit{Security, Reliability and Trust}\\\textit{University of Luxembourg}\\
Luxembourg, Luxembourg \\
radu.state@uni.lu}
}
\end{comment}
%%% END OLD AUTHORS BLOCK %%%

\author{
\IEEEauthorblockN{
Lucian Trestioreanu\IEEEauthorrefmark{1},
Flaviene Scheidt\IEEEauthorrefmark{1},
Wazen Shbair\IEEEauthorrefmark{1},
Jerome Francois\IEEEauthorrefmark{1},
Damien Magoni\IEEEauthorrefmark{2},
and
Radu State\IEEEauthorrefmark{1}
}
 \IEEEauthorblockA{\IEEEauthorrefmark{1} University of Luxembourg, SnT, 29, Avenue J.F Kennedy, L-1855 Luxembourg\\
 Email:\{lucian.trestioreanu, flaviene.scheidt, wazen.shbair, jerome.francois, radu.state\}@uni.lu\\
 }
  \IEEEauthorblockA{\IEEEauthorrefmark{2} University of Bordeaux, LaBRI - CNRS, 351, Cours de la Liberation, F-33405 Talence, France\\
 Email: damien.magoni@u-bordeaux.fr\\
 }
 }

%\author{Anonymous Submission}
%\{lucian.trestioreanu\}@uni.lu}
\maketitle

% !TEX root = main.tex

\begin{abstract} 

With the large increase in the adoption of blockchain technologies, their underlying peer-to-peer networks must also scale with the demand. In this context, previous works highlighted the importance of ensuring efficient and resilient communication for the underlying consensus and replication mechanisms. However, they were mainly focused on mainstream, Proof-of-Work-based Distributed Ledger Technologies like Bitcoin or Ethereum.

In this paper, the problem is investigated in the context of consensus-validation based blockchains, like the XRP Ledger. The latter relies on a Federated Byzantine Agreement (FBA) consensus mechanism which is proven to have a good scalability in regards to transaction throughput. However, it is known that significant increases in the size of the XRP Ledger network would be challenging to achieve. The main reason is the flooding mechanism used to disseminate the messages related to the consensus protocol, which creates many duplicates in the network. Squelching is a recent solution proposed for limiting this duplication, however, it was never evaluated quantitatively in real-life scenarios involving the XRPL production network. In this paper, our aim is to assess this mechanism using a real-life controllable testbed and the XRPL production network, to assess its benefit and compare it to alternative solutions relying on Named Data Networking and on a gossip-based approach.

\begin{comment}
    
Towards closing this gap, we present here our journey to provide a solution for increasing the communication efficiency for consensus-validation based blockchains like the XRP Ledger~\cite{DBLP:journals/corr/abs-1802-07242}. More specifically, we present the experimentation with, and the evaluation of, a novel message optimisation technique called \textit{Squelching}. We experiment by sending XRP Ledger's consensus messages over a dedicated testbed deployed on \textit{Grid 5000} and show in practice that \textit{Squelching} is indeed able to dramatically lower the total number of messages that travel on the XRP Ledger peer-to-peer network. The conclusion is drawn from a top-down view, after discussing and comparing \textit{Squelching} with two other overlay-based solutions that we previously identified, implemented, and evaluated: one based on \textit{Named Data Networking} and the second on \textit{Gossipsub}. 

However, this technology is still relatively new and the task of scaling has come with its own challenges. As such, the scalability of blockchain networks is an active problem in the community and also in research. 

\end{comment}

\end{abstract} 

%thttps://www.overleaf.com/project/5e419f39864fb200015c5139

% !TEX root = main.tex

\section{Introduction}
\label{sec:intro}

Distributed Ledger Technology (DLT) is relatively new and still evolving. Its development was fostered by a diverse and enthusiastic community that sometimes forgot lessons from the past related to efficiency, resilience, and security of communication. Many blockchains rely on \textit{flooding} as a straightforward solution for addressing the one-to-many and many-to-many communication specifics of DLT, which leads to scalability limitations. In this context, scalability concerns either being able to add more nodes to the network (\textit{Node}-wise scalability), or the ability to process more transactions per second (TPS) (\textit{Throughput}-wise scalability).
%DONE (Added Falcon, Bloxroute, Fibre projects) 
%\JF{New proposition for structure below. We need a paper citation for BTC message transmission optimization XXXX below}
\textit{Throughput}-wise scalability can be achieved on-chain (sharding), off-chain (payment channels, sidechains), or otherwise e.g. specific mechanisms like XRP Ledger's (XRPL) consensus,%. Unlike Proof-of-Work (PoW) and Proof-of-Stake-based blockchains which rely on computational power or stakes respectively, XRPL is a Byzantine Fault Tolerant (BFT)-based blockchain 
~which uses votes of groups of trusted validators to achieve consensus. 

Generally, PoW blockchains can easily increase their number of nodes, but face challenges to increasing throughput, while Byzantine Fault Tolerant (BFT) blockchains offer higher throughput but can not easily scale node-wise~\cite{Vukolic-scalability-PoWnodes-BFTTx}. For instance, XRPL %can achieve 1500 TPS (with 3000 TPS expected soon~\cite{DBLP:journals/corr/abs-1802-07242,RPCA-old}), but 
~relies on a peer-to-peer (P2P) flooding mechanism leading to a multitude of redundant messages being circulated. %With a network size of around 1000 nodes, this already has a negative impact on scalability~\cite{Tsipenyuk-valnumber}.

Previous work highlighted how latency and bandwidth of the underlying communication from physical channels to protocols can limit the scalability~\cite{latency-bw-blckchain-scalability,latency-blockchain-sclability}. In this perspective, the community effort was mainly focused on mainstream blockchains like Ethereum (ETH)~\cite{gossipsub-paper, NDN-Ethereum} or Bitcoin (BTC), both being Proof-of-Work type (PoW) (recently, ETH switched to \textit{Proof-of-Stake}). Projects like \textit{Fibre}\cite{fibre}, \textit{Falcon}\cite{falconSpringer} or \textit{bloXroute}\cite{Klarman2018bloXrouteA}, aimed to improve BTC transaction rate by speeding up block propagation. \textit{Node}-wise scalability can be achieved for example by efficient message transmission options, but was not well explored in the case of consensus-validation based blockchains~\cite{Vukolic-scalability-PoWnodes-BFTTx}. Therefore, this paper focuses on the \textit{Node}-wise scalability of BFT-based blockchains with XRPL as an illustrative example. %, since this is not well-explored yet. % DM already said above

To mitigate the overhead of message flooding, \textit{Squelching} is a recently proposed dissemination protocol, that decreases the number of messages on the XRPL P2P network by reducing the number of duplicates. The main principle resides in a careful selection of peers from which to receive messages, rather than receiving messages from all possible peers.

This work evaluates to what extent \textit{Squelching} can improve the performance of intra-ledger communication on XRPL, (\textit{i.e.} decrease the number of consensus-related messages), thus %leading towards 
reducing the computational overhead, and ultimately improving XRPL's \textit{Node}-wise scalability. For evaluation purposes, two types of experiments are performed: (1) baseline experiments on the actual XRPL network measure the impact of  flooding on the computational overhead of a node (without Squelching) and (2) experiments in a controlled distributed environment to evaluate the benefit of Squelching. Together, these experiments allow assessing how Squelching improves P2P connectivity. This paper contributes to filling a research gap regarding the efficiency of the underlying communication supporting the consensus protocols of BFT blockchains like XRPL. 

Our contributions are four-fold: i) Highlight how the current flooding mechanism of XRPL contributes to limiting its \textit{Node}-wise scalability; ii) Define a measurement method to evaluate the impact of Squelching in regards to a baseline version; iii) Apply the aforementioned method to assess Squelching quantitatively; iv) Discuss the results in the context of other proposed alternatives, namely XRP-NDN~\cite{XRP-NDN-overlay} and GossipSub~\cite{gossipsub-Flaviene}.

The structure of the paper is as follows: Section~\ref{sec:perf} introduces background on XRPL and refines the problem. Section~\ref{sec:design} presents \textit{Squelching.} The evaluation method and results obtained are shown in Section~\ref{sec:results}. In Section~\ref{sec:discussion}, alternative solutions are discussed; Section~\ref{sec:relwork} provides a broader overview of related work, and Section~\ref{sec:conclusion} concludes.

%\section{background and related work}
%\label{relatedwork}
% !TEX root = main.tex

\section{Problem refinement}
\label{sec:perf}

\subsection{Background on the XRP Ledger}

Leslie Lamport previously showed that in a synchronous environment, consensus can be achieved if at most \(n\) out of \(3n+1\) parties involved are dishonest~\cite{lamport1982-BFT}. The \textit{Practical Byzantine Fault Tolerance}~\cite{PBFT} algorithm makes this practicable in asynchronous environments like Internet, paving the way to consensus-validation based DLTs. XRPL implements a variant of BFT consensus called Federated Byzantine Agreement (FBA) Consensus~\cite{XRPL-FBA-consensus}: named the \textit{XRP Ledger Consensus Protocol} (XRP LCP), it improves transaction (TX) throughput while maintaining  security against Byzantine failures.
%The \textit{XRP Ledger Consensus Protocol} (XRPLCP)~\cite{DBLP:journals/corr/abs-1802-07242},\cite{RPCA-old} belongs to the family of mechanisms known as \textit{Byzantine Fault Tolerant} (BFT) consensus algorithms. 
Although the baseline algorithm supposes all nodes to agree on the list of participants and to process all transactions to reach a consensus, FBA introduces the concept of quorum slices: a node only needs to trust a subset of the other nodes, to take its own decision about the TX to be validated. %Because nodes can decide themselves which other nodes to trust, multiple quorum slices can form in the network. For a healthy network, the quorum slices must overlap to some degree.%, as shown in Figure~\ref{fig:FBA-quorum-slice-overlap}. 
%~This means some nodes are trusted in multiple quorum slices, thus ensuring the dissemination of information. The slices must reach a decision among themselves, and after this, a final decision can be taken.

%		 \begin{figure}[ht!]
%   			\begin{minipage}{0.5\textwidth}
%   			\centering
%       		\includegraphics[width=0.6\textwidth]{images/Th-blockchain-XRPL-UNLslice-overlap.pdf}
%			\caption[FBA quorum slices]{FBA quorum slices.}
%            \label{fig:FBA-quorum-slice-overlap}
%        	\end{minipage}
%		\end{figure}

\begin{figure}[ht!]
\centering
%\vspace{13pt}
\includegraphics[width=0.4\textwidth]{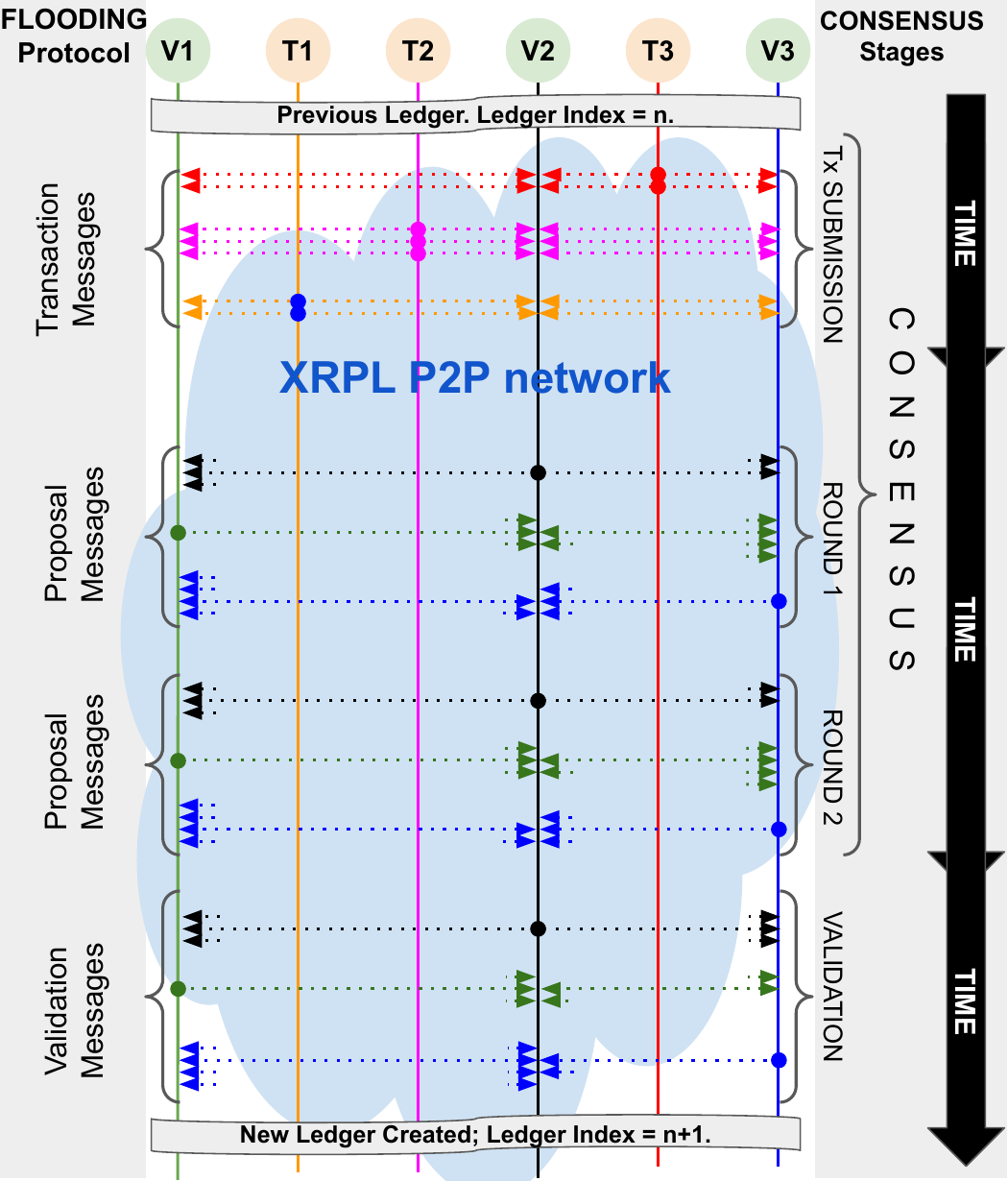}
\caption{The Ledger building process on XRPL.}
\label{fig:xrpl_consensus}
\end{figure}

On XRPL, there are two main node types: \textit{trackers} in charge of processing transactions, and \textit{validators} used for consensus voting. %Each voter (validator) selects a Unique Node List (UNL) of \textit{trusted} voters to collaborate with, meaning they are \textit{trusted} not to collude among themselves to defraud the respective validator. 
Figure~\ref{fig:xrpl_consensus} illustrates the ledger creation. A new ledger is created from the previous one by applying a new set of Tx's, a process involving three main phases:

\textit{1. Transaction submission:} New transactions can be submitted through flooding at any time, but for a certain ledger, there is a time window in which the new transactions can be accepted. The network waits for a certain time for new \textit{transactions} to be included in the current ledger: during this time, new transactions submitted from the tracker nodes T1, T2, T3, are flooded as \textit{transaction} messages through the XRPL network and will be received multiple times by the validators V1, V2, V3. New transactions that missed the time window are stored and will be processed in the next ledger. While validators can technically also submit transactions, this is a discouraged practice for security and performance reasons.

\textit{2. Consensus:} During multiple consensus rounds, validators exchange \textit{proposal} messages to agree on the transaction set to be included. Due to the flooding mechanism that is used, the network nodes produce and propagate duplicate messages. 

\textit{3. Validation}: Validators may produce inconsistent ledger versions for the same ledger index. Therefore, in this phase, validators exchange \textit{validation} messages to agree on the next ledger to be created from all candidates for the given index.

\subsection{XRPL topology}
\label{sec:topo}

XRPL uses flooding for the dissemination of transaction, proposal, and validation messages. This is effective to explore every path and reach every node, but inefficient because it also forwards a significant number of duplicates; but this also depends on the underlying topology. As a preliminary study, we scanned and analyzed XRPL's topology as of 2021 with Nem~\cite{magoni:hal-00344484}: %Results are summarised in Table~\ref{tab:XRPL-topo-analysis}.
it consisted of 892 nodes and 9197 edges; 152 nodes were \textit{validators}. The average distance between a pair of nodes was 2.37, \textit{topology diameter} was 5, and average degree 20.62. %A giant bi-connected component of 867 nodes was identified, which was composed of 9172 edges; mean degree of the nodes was 21, while max. degree was 296, with multiple similarly high-degree nodes. \textit{Mean distance} was 2.34, the \textit{median distance} 2.37, with a \textit{diameter} of 4. %An in-depth analysis can be found in~\cite{topo-xrpL}. 
Thus on one hand, XRPL topology is dense, highly connected, and with a low diameter. On the other hand, it is known that a dense network is more resilient to node and edge failures but its communication performance is affected by the high number of duplicates incurred when flooding is used, as the number of messages is proportional to number of edges~\cite{perigee}. %It is worth noticing that each message including duplicates, will go through multiple processing steps including parsing, signature verification, etc.

\subsection{Objective}

Although XRPL \textit{throughput} can reach 1500 TPS, scaling it \textit{Node}-wise can be challenging due to the inefficient message flooding used and the dense network structure. Also as previously shown in~\cite{Tsipenyuk-valnumber}, the number of XRPL \textit{proposal} and \textit{validation} messages represents 72\% of all messages, so it is worth optimizing them. To enhance XRPL scalability by improving communication efficiency when disseminating these particular messages,~\textit{Squelching} was recently proposed. However, no in-depth assessments were performed. In this paper, our goal is to benchmark this solution, evaluate its potential benefit and compare the results with our previous assessments of other solutions based on NDN \cite{XRP-NDN-overlay} and Gossipsub \cite{gossipsub-Flaviene}.

%focusing on  optimising the dissemination of these messages only can vastly improve the communication efficiency on XRPL. Our a and this is the goal of the present work.

%\input{background}

% !TEX root = main.tex

\section{Squelching}
\label{sec:design}

\subsection{Overview}

The \textit{Squelching} protocol was designed to optimize message relaying in the XRPL network. Its main goal is to reduce bandwidth consumption, CPU and memory load, and improve \textit{Node}-wise scalability. Assuming a given validator, each node selects a subset of its peers to relay messages created and flooded by the given remote validator. In parallel, it sends a \textit{squelch} message to the rest of its peers to \textit{squelch} the connection for a given time, i.e., stop relaying messages to it. Hence, it is an active solution where a node assumes other nodes to behave compliant to the requests they receive. By reducing the number of relaying connections (network edges) involved in the flooding process, and in addition by selecting lower latency connections, this protocol reduces the number of exchanged messages and so, the load on the network and the processing hosts. As a result, performance is expected to increase on multiple facets (lower CPU and bandwidth usage, lower message latency). We aim at quantitatively evaluate the achievable performance gain on production hosts.

\begin{figure}[htb!]
\begin{center}
    \includegraphics[width=0.7\linewidth]{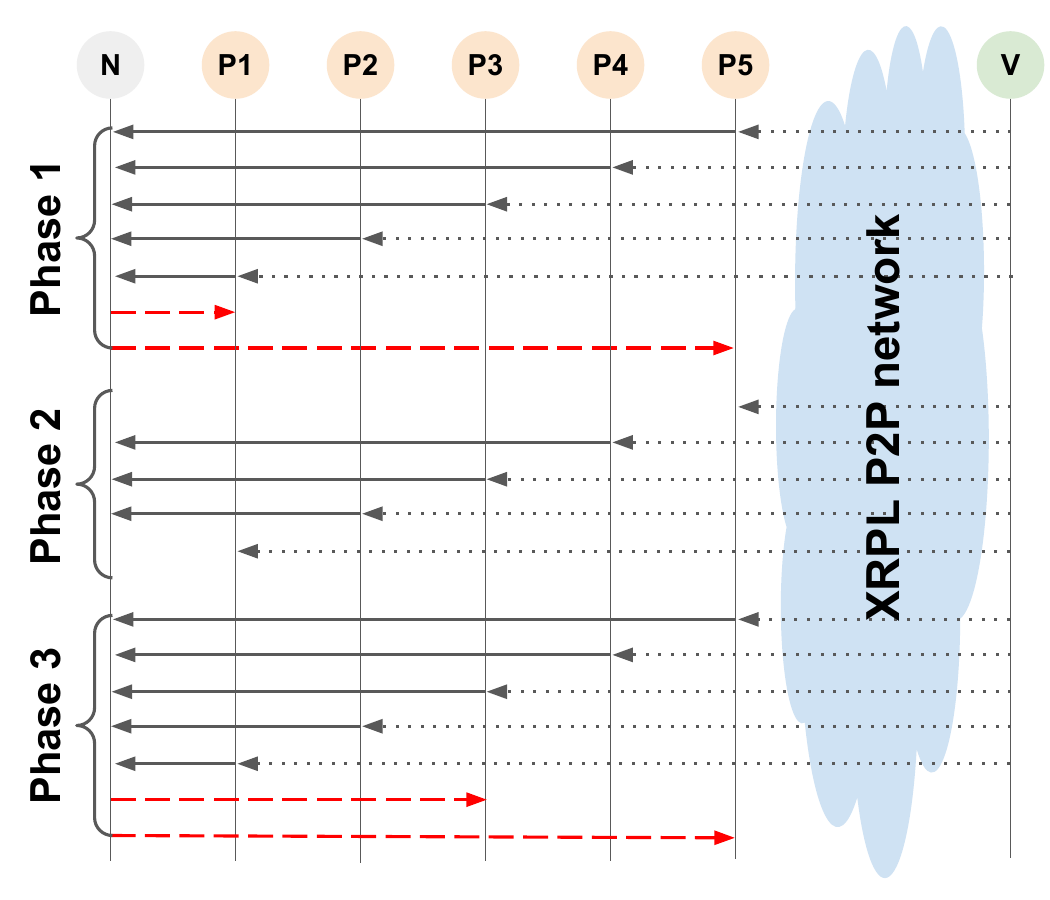}
    \caption{Squelching sequence diagram.}
    \label{fig:squelching-workflow}
\end{center}
\end{figure}

Figure~\ref{fig:squelching-workflow} exemplifies the squelching protocol~\cite{Tsipenyuk-valnumber}, where a node switches between phase 2 and 3 once initialization is done (phase 1):

\textit{Phase 1:} (1) Remote \textit{validator V} creates and floods validation and proposal messages on the XRPL network, which reach \textit{P1-P5} (either trackers or validators) via possibly different routes (dotted arrows). (2) Nodes \textit{P1-P5} are direct peers (next-hop connections) of node \textit{N} and relay the messages from \textit{validator} \textit{V} to node \textit{N} (black arrows). (3) As a result, the node \textit{N} receives each message five times. In this example, it determines that messages from nodes \textit{P2}, \textit{P3}, and \textit{P4} arrive faster. The latency of the connections from these peers is thus lower. The node \textit{N} selects these peers to relay the messages from the \textit{validator} \textit{V}, and sends \textit{squelch} messages to peers \textit{P1} and \textit{P5} (dashed red arrows). 

 \textit{Phase 2:} \textit{P1} and \textit{P5} continue to receive messages from \textit{validator V}, but only \textit{P2}, \textit{P3}, \textit{P4} relay them to node \textit{N}.

  \textit{Phase 3:} (1) After some time, \textit{P1} and \textit{P5} un-squelch themselves and start relaying again messages from validator V to node \textit{N}. (2) Node \textit{N} determines now that peers \textit{P1}, \textit{P2}, and \textit{P4} are better candidates, and sends \textit{squelch} messages to nodes \textit{P3} and \textit{P5}.

\subsection{Slot management}

For explanation clarity, \textit{uplinks} are the direct peers from which an XRPL node receives messages, while \textit{downlinks} are the direct peers to which the given node sends messages to. So a node relays messages from its \textit{uplinks} to \textit{downlinks}.

\textit{Squelching} introduces the concept of \textit{slots} and \textit{squelches} to manage message relaying among nodes. 
%This protocol allows each server to maintain a set of \textit{uplinks} for each slot, optimizing the relaying process. 
Each node creates a \textit{slot} for each validator it chooses to relay messages for. For each slot (validator), a list of uplinks and downlinks is maintained. When a node (tracker or validator) receives a \textit{proposal} or \textit{validation} message from one of its uplinks, it extracts the originating validator. The node then checks the corresponding slot, and if enough copies have been already received, a \textit{squelch} message is sent back to the peer that just relayed this message. The squelch message asks the peer to stop forwarding further messages created by the respective originating validator.% to the relaying server over the current connection.
~When receiving the \textit{squelch} message, a peer removes the sender of the message from the \textit{downlinks} for the corresponding slot (originating validator). Hence, each node keeps track of its downlinks towards which the messages must be relayed for every unique validator.

\textit{Squelches} have a time limit allowing for network topology changes. After some time a squelch expires and the respective peer can be considered again by the node as candidate relay for the given originating validator. If a node loses an \textit{uplink} it can replace it by sending \textit{unsquelch} messages to peers. This allows to maintain enough connected peers to ensure security.  

%Servers can also send an "unsquelch" message to recover their uplink if they lose it.

%REMOVED FOR MINI_CONFERECE
%\subsection{Implementation}

%The protocol introduces a new XRPL message type for instructing peers to \textit{squelch} or \textit{unsquelch} messages from specific validators. Feature negotiation is done via HTTP headers to determine whether a node supports the feature. So, the protocol can be implemented without the need for a network-wide upgrade. Obviously, if a node is not compatible, it cannot interpret the \textit{squelch messages} and continues forwarding all messages. Hence, consensus can still be achieved. While the number of peers to squelch could be a configurable parameter, XRPL software sets this number to five. 

\begin{comment}

To evaluate the overlays proposed we modified the code of the XRPL nodes to be able to redirect the validation mesages through the NDN overlay or through Gossipsub. To achieve this we used the architecture presented in Figure~\ref{fig:xrpndn-architecture-grpc}.

\begin{figure}[htb!]
\begin{center}
    \includegraphics[width=0.4\textwidth]{figures/Squelch/TH-XRP-NDN_gRPC-architecture.pdf}
    \caption[General architecture of XRP-NDN Overlay]{General architecture of XRP-NDN Overlay.}
    \label{fig:xrpndn-architecture-grpc}
\end{center}
\end{figure}

\end{comment}

% !TEX root = main.tex

\section{Evaluation}
\label{sec:results}

\subsection{Method and metrics}

In this section, the main objective is to measure the impact of squelching in comparison with a baseline implementation without squelching, i.e., using flooding.

After a \textit{first set} of experiments on XRPL MainNet we observed that mostly CPU usage is significantly impacted by the number of messages to handle. These experiments focused on the CPU usage by an XRPL node without squelching but with a varying number of peers, since the number of flooded messages received directly depends on the latter. So, on a node connected to MainNet, we measured as a function of number of peers: average CPU usage, number of messages received, and number of messages sent.
%Considering both bytes and the number of messages is relevant as the processing a of message directly depends directly on its content and so on its size, for example when it must be decrypted. 
This enabled us to quantify the CPU overhead and messages to be processed due to an increased number of peers, and infer a regression model.

In the \textit{second set} of experiments, to better control the experimental parameters, we deploy a small-scale XRPL network on a controlled and configurable HPC facility; squelching is applied, and its impact is monitored in terms of number of messages sent and received. Thanks to regression modeling, we will then extrapolate the number of messages saved in the Mainnet XRPL network and ultimately the number of free slots for additional peers to connect to. The baseline used for evaluation was an unmodified version of XRPL (XRPL v1.6), which was compared with a version~\cite{squelch-git} implementing the Squelching mechanism (XRPL v1.7).

%\begin{figure}[ht!]
%    \centering
%     \includegraphics[width=0.99\linewidth]{images/chart (29).pdf}      
%        \captionsetup{justification=centering}
%        \caption{Mean CPU usage according to the number of peers.}
%        \label{fig:CPU}
%\end{figure}

\begin{figure}[ht!]
      %\subfloat[Messages IN]{
       \subfloat{\includegraphics[width=0.47\columnwidth]{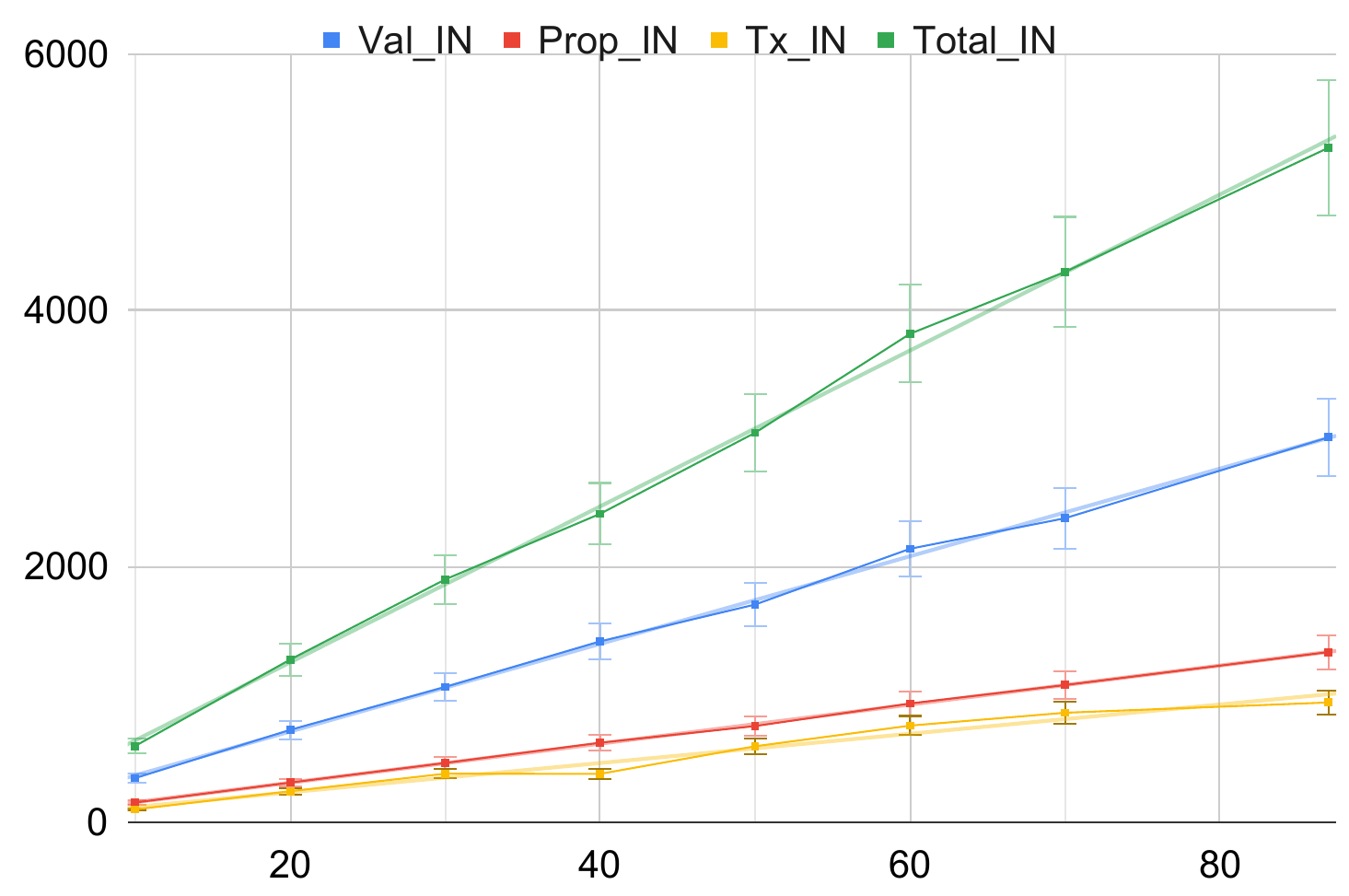}
    \label{fig: lin-msgs in vs peers}
    }  
    \subfloat{\includegraphics[width=0.47\columnwidth]{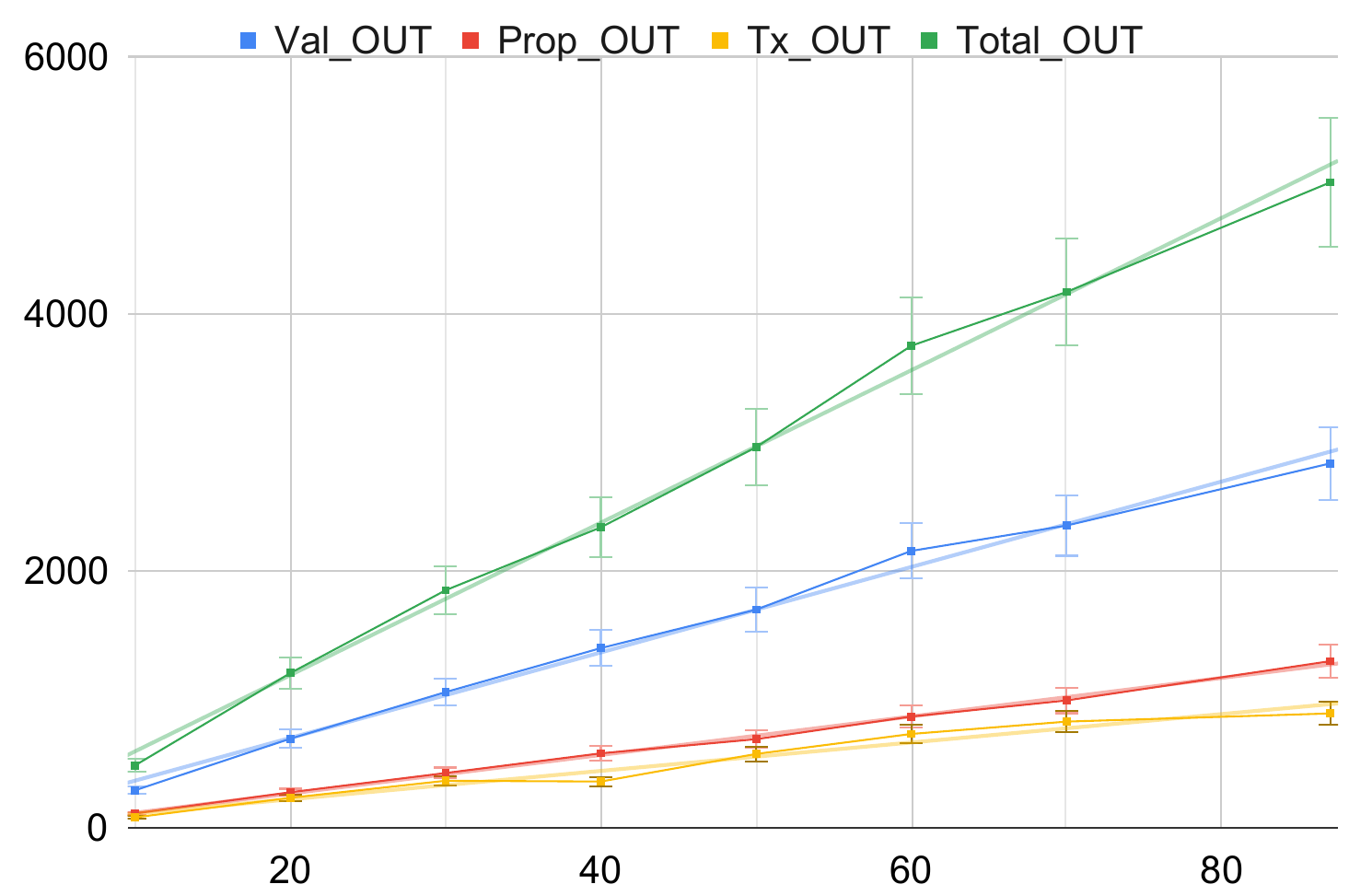}
    \label{fig: lin-msgs out vs peers}
    }
    \captionsetup{justification=centering}
      \caption{IN/OUT number of msgs. \textit{(y)} \textit{vs} number of peers \textit{(x)}, \\ (Val: validation, Prop: proposal, Tx: transaction).}
  \label{fig:msgs vs peers}
\end{figure}

\begin{table}[ht!]

\captionsetup{justification=centering}
\caption{Mean CPU usage according to number of peers.}

\centering
\resizebox{\columnwidth}{!}{%
\begin{tabular}{l|ccccccc}
\textbf{Peers} & 10     & 20     & 30     & 40     & 50     & 60     & 70     \\ \hline
\textbf{CPU}   & 16.929 & 18.651 & 18.694 & 20.645 & 22.405 & 22.924 & 23.825
\end{tabular}
}

        \label{tab:CPU}

\end{table}

\begin{comment}

\begin{figure*}[ht!]
    \centering
    \captionsetup[subfigure]{justification=centering}
    \begin{minipage}{0.33\textwidth}
     \includegraphics[width=0.99\textwidth]{images/Linear-dep-msgs-peers/chart (25).pdf}      
        \captionsetup{justification=centering}
        \caption{Mean CPU usage versus \\ number of peers.}
        \label{fig:CPU}
    \end{minipage}
\begin{minipage}{0.66\textwidth}
       %\subfloat[Messages IN]{
       \subfloat{ \includegraphics[width=0.49\textwidth]{images/Linear-dep-msgs-peers/chart (27).pdf}
    \label{fig: lin-msgs in vs peers}
    }  
%\subfloat[Messages OUT]{
\subfloat{        \includegraphics[width=0.49\textwidth]{images/Linear-dep-msgs-peers/chart (28).pdf}
    \label{fig: lin-msgs out vs peers}
    }
    \captionsetup{justification=centering}
      \caption{Number of messages IN (left) and OUT (right), per second and per type (Val: validation, Prop: proposal, Tx: transaction).}
  \label{fig:msgs vs peers}
\end{minipage}
\end{figure*}
\end{comment}

\subsection{Baseline experiments (first set, XRPL MainNet)}

To perform the baseline experiments, a single node run on Ubuntu 22.04 with 16Gb RAM and 4 Intel Xeon E5-4650 v4 @2.2GHz is connected to the XRPL MainNet. Statistics are gathered with the RippledMon tool~\cite{rippledmon}.

We configured the node to connect to an increasing number of peers over time, and collected average CPU usage and the number of messages sent and received. Results given in Table~\ref{tab:CPU} and Figure~\ref{fig:msgs vs peers} show a linear dependency between the number of peers and number of messages. 
As expected, total number of messages received and sent increases with the number of peers, as the flooding mechanism is exploiting the new connections to send redundant messages. %There are no significant differences between the direction of messages, since for each new message received, our node also sends duplicatas. 
~Due to the BFT-based validation mechanism, the number of exchanged messages varies according to the type of message, with \textit{validations} being the most frequent. As such, Squelching could be fine-tuned for this type of message. This will still improve \textit{Node}-wise scalability while enforcing higher robustness for other message types. 

Due to a higher number of messages to be processed %(parsing, signature check, etc)
when the number of peers increases from 10 to 70, CPU usage also increases gradually approximately by 35\%, as in ~Table~\ref{tab:CPU}.

\begin{comment}

\begin{figure*}[t]
  \centering
  \subfloat[Total number of Messages IN]{
        \includegraphics[width=0.47\textwidth]{images/G5000/Th-G5000-results-annotated (1).pdf}
    \label{fig: xrpl vs squelch msgs in}
  }
  \subfloat[Total number of Messages OUT]{
        \includegraphics[width=0.47\textwidth]{images/G5000/G5000 Messages OUT.pdf}
    \label{fig: xrpl vs squelch msgs out}
  }

  \subfloat[Total number of Bytes IN]{
        \includegraphics[width=0.47\textwidth]{images/G5000/G5000 Bytes IN.pdf}
    \label{fig: xrpl vs squelch msgs in}
  }
\subfloat[Total number of Bytes OUT]{
        \includegraphics[width=0.47\textwidth]{images/G5000/chart (21).pdf}
    \label{fig: xrpl vs squelch msgs in}
    }
  
  \caption{Cumulative number of messages: unmodified XRPL vs XRPL with squelching}
  \label{fig: xrpl vs squelch}
\end{figure*}
\end{comment}

\subsection{Performance model}
%%%% Regeression model

%COmment to submit
%\JF{Modify the first regression below}

To assess the \textit{Node}-wise scalability, we model the relationships between the resources used in terms of CPU usage, messages processed, and number of peers of an XRPL node. This allows to quantify the resources used by a node. Thanks to a linear dependency, simple regression allows to model CPU use ($cpu$) with respect to the number of peers the node is connected to ($peers)$:
\begin{equation}
    \begin{aligned}
%    cpu=\beta_0+\beta_1 msgs\_in + \beta_2 msgs\_out\\
%with\ \beta_0=14.5837, \beta_1=0.0158, \beta_2=-0.0137
cpu=\beta_0+\beta_1 peers\\
with\ \beta_0=15.8754, \beta_1=0.1177
\end{aligned}
\label{eq:regressioncpu}
\end{equation}

Messages are considered independently of their type. We do not distinguish CPU used by each type. Similarly, the total number of messages ($msgs$) is linearly dependent to number of peers ($peers$):

\begin{equation}
    \begin{aligned}
msgs = \alpha_0+\alpha_1 peers \\
with\ \alpha_0=-75.0943, \alpha_1=123.6365
\end{aligned}
\label{eq:regressionmsgs}
\end{equation}

Because of a good linearity of the underlying data, the regression models fit with a high coefficient of determination, $R^2>0.96$ for both equations~(\ref{eq:regressioncpu}) and (\ref{eq:regressionmsgs}).

\subsection{Setup for squelching experiments (second set, Grid5000)}

To assess Squelching, we deployed our own XRPL network to fully control the nodes.
%To perform the measurement in real-life conditions where nodes are spread among different geographies, 
We built the testbed on Grid5000 (G5K)~\cite{G5K},\cite{grid5000}, a large-scale HPC platform with interconnected sites in France and Luxembourg, featuring a large amount of resources: 15000 cores, 800 nodes with bare-metal access, and 10Gbps Ethernet links. To deploy the XRPL network on G5K and perform the evaluation, we used our previous work, BlockZoom~\cite{block-zoom},\cite{blockzoom} offering a reproducible environment for DLT experimentation.
%        \begin{figure}[ht!]
%   			\begin{minipage}{0.5\textwidth}
%   			\centering
%       		\includegraphics[width=1\textwidth]{images/G5000/Th-G5000-testbed.pdf}
%			\caption[XRP network monitoring testbed deployed on G5K.]{XRP network monitoring testbed deployed on G5K.}
%            \label{fig: G5000 testbed}
%        	\end{minipage}
%		\end{figure}

The testbed was composed of 15 XRPL nodes evenly spread over five \textit{G5K} locations (Luxembourg, Rennes, Nantes, Sophia, Lyon), meaning three nodes per site.%, to simulate the real-life conditions where XRPL nodes are spread over different locations, with different latencies. Because latencies impact message propagation performance, we prefer this setup instead of others with limited latency. For instance, while latencies can be added in other ways, e.g., using Mininet, this is still not realistic. Furthermore, 
~Alternative solutions like Mininet come with their own limitations and challenges, for example, the requirements to run many XRPL nodes on the same machine. On the other hand, G5K is a research-oriented platform providing monitoring tools to access precisely the performance of a running algorithm/software and hardware on the platform. Moreover, G5K supports experiment reproducibility which is crucial for scientific analysis.

%The consensus quorum was set to six (in real life not all nodes on the production XRPL network are validators). 
The nodes boot with no transactions being generated until a predefined time. Once this time expires, 1000 transactions per site were sent in parallel. 

%Done - added details above
%\JF{WHy this setup? Is this consistent with the XRPL real topology you present before, for example in terms of structure or density? We somehow need to show we have a smaller scale envrionment but with similar properties (in particular because say in real life...)}

\begin{figure}[t]
  \centering
  \subfloat[Total number of Messages IN]{
        \includegraphics[width=0.47\columnwidth]{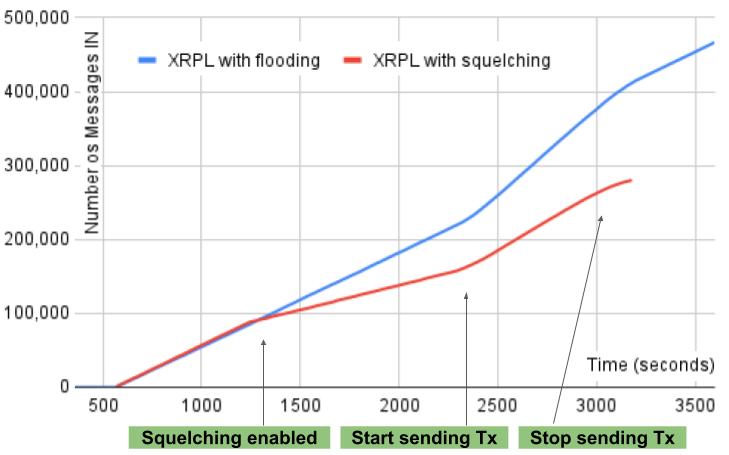}
    \label{fig: xrpl vs squelch msgs in}
  }
  \subfloat[Total number of Messages OUT]{
        \includegraphics[width=0.47\columnwidth]{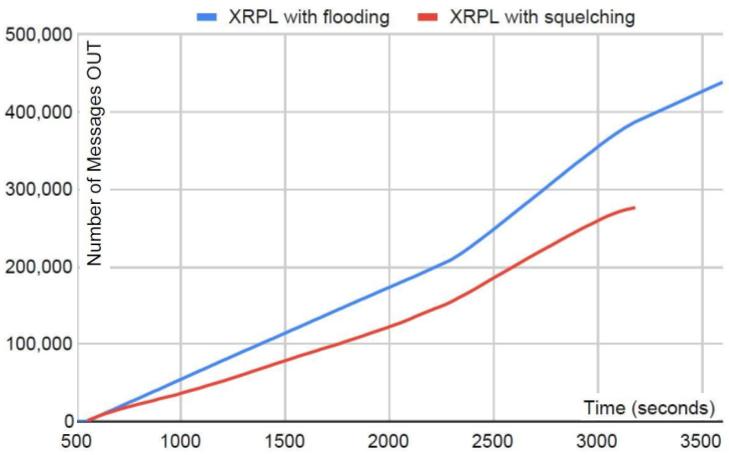}
    \label{fig: xrpl vs squelch msgs out}
  }

\begin{comment}
    
  \subfloat[Total number of Bytes IN]{
        \includegraphics[width=0.47\textwidth]{images/G5000/G5000 Bytes IN.pdf}
    \label{fig: xrpl vs squelch msgs in}
  }
\subfloat[Total number of Bytes OUT]{
        \includegraphics[width=0.47\textwidth]{images/G5000/chart (21).pdf}
    \label{fig: xrpl vs squelch msgs in}
    }
\end{comment}
  
  \caption{Total number of messages, \textit{unmodified} vs. \textit{squelching}.}
  \label{fig: xrpl vs squelch}
\end{figure}

\begin{table}[]
\centering
\caption{Squelch versus Flooding results.} 
\label{tab:Squelch-results}
\begin{tabular}{l|l}
FLOOD - average of total messages / second   & 297.633 \\ \hline
SQUELCH - average of total messages / second & 211.602 \\ \hline
SQUELCH / FLOOD (\%)                         & 71.094  \\ \hline
SQUELCHING saves over FLOOD (\%)             & 28.905 
\end{tabular}
\end{table}

\subsection{Overall Results}

The comparative results of G5K experiments (second set) are presented in Figure~\ref{fig: xrpl vs squelch} and Table~\ref{tab:Squelch-results}. Figure~\ref{fig: xrpl vs squelch}\subref{fig: xrpl vs squelch msgs in} is annotated with the stages of the experiment for the sake of clarity, \textit{i.e.} when squelching is enabled and when the node starts/stops sending \textit{tx} messages.

As they are cumulative functions, the difference in the number of messages between flooding and squelching increases over time. The same observation applies to received messages at a slightly lower scale. When the transactions start to be sent, the number of messages logically increases and the observed differences are still valid.

From Table~\ref{tab:Squelch-results} we see that \textit{Squelching} saves 28.9\% messages per second on average over \textit{Flooding}. It must be noted that on the G5K experiments, at 15 peers, the number of messages is much lower than its 15-peers equivalent on Mainnet (Figure~\ref{fig:msgs vs peers}). This is due to the much larger scale of Mainnet, which generates proportionally more messages. However, this is not an impediment, because what we are interested in is the last row in Table~\ref{tab:Squelch-results}, \textit{i.e.}, the average savings offered by Squelching \emph{vs} Flooding (\%). The latter can be translated directly to the Mainnet experiment thanks to our regression models.

\begin{figure}[ht!]
    \centering
     \includegraphics[width=0.99\linewidth]{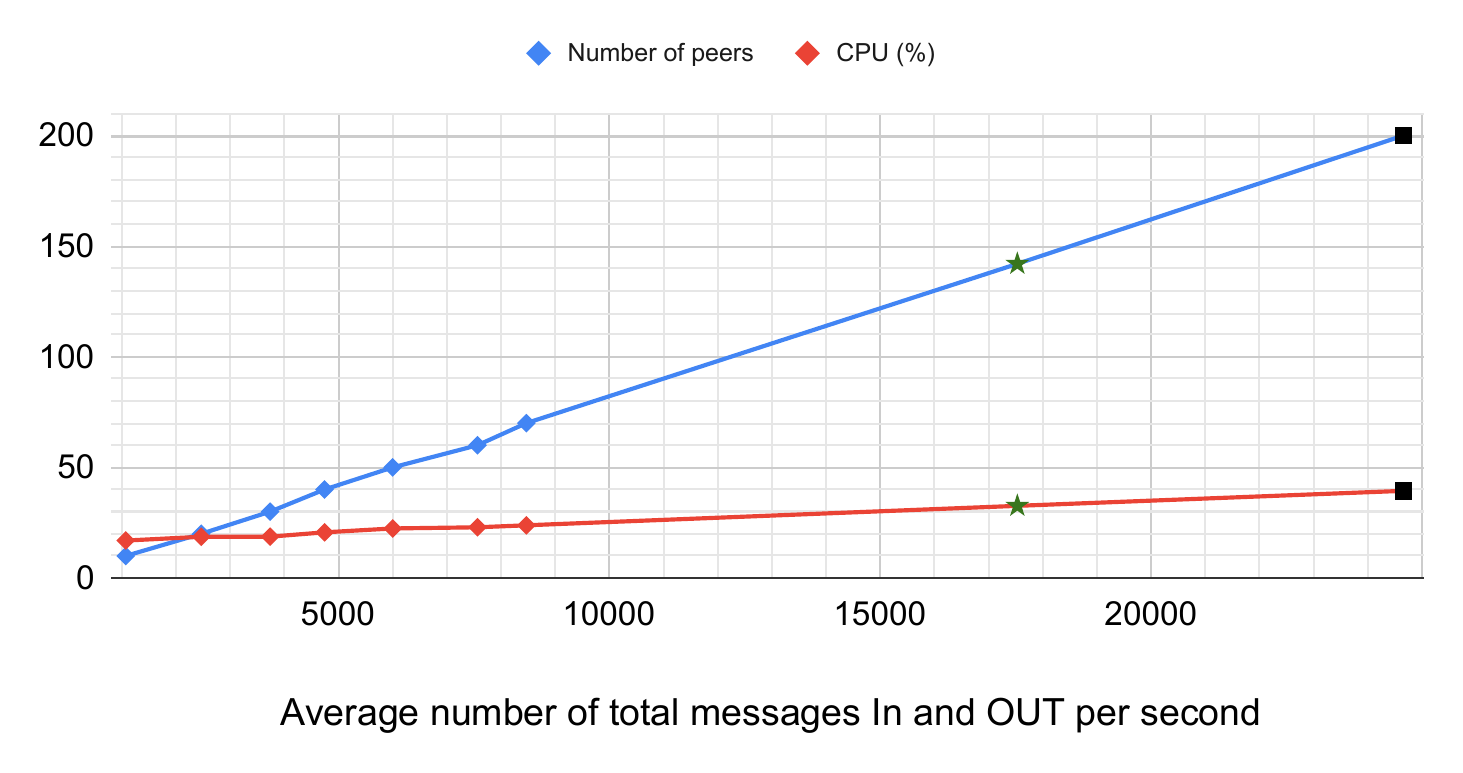}      
        \captionsetup{justification=centering}
        \caption{Extrapolation of \textit{number of peers} and \textit{CPU(\%)}.}
        \label{fig:gains-regression}
\end{figure}

Table~\ref{tab:Squelch-gains} describes how we computed the potential gain of an XRPL hub node with 200 peers on Mainnet if \textit{Squelching} would be applied. The baseline measurements are reported as points \#1 to \#7, reminding that they were measured experimentally on Mainnet. Point \#9 (black squares in Figure~\ref{fig:gains-regression}) is obtained by the linear regression models of equations (\ref{eq:regressioncpu}) and (\ref{eq:regressionmsgs}), by rearranging the terms as follows:
\begin{equation}
    \begin{aligned}
    CPU = 0.1177*Peers + 15.8754 \\
    Peers = 0.0081 * Messages + 0.6074\\
\end{aligned}
\label{eq:regressionCPU-msgs}
\end{equation}

Then, knowing that squelching could reduce the amount of messages by 28.9\%, we compute the number of peers and CPU for 28.9\% less messages than Point \#9, i.e., for 17527 messages. For this value, we compute the corresponding number of peers and CPU (Point \#8, green stars in Figure~\ref{fig:gains-regression}) with our performance model (equations (\ref{eq:regressioncpu}) and (\ref{eq:regressionmsgs}) again).

The result shows that a node connected to XRPL MainNet with 200 peers could save 17\% CPU with \textit{Squelching} and free up 58 peer slots. This actually enhances overall connectivity as a single node has 29\% additional peers to connect to.

\begin{table}[]
\centering
\caption{Gains obtained through Squelching.} 
\label{tab:Squelch-gains}
\begin{tabular}{ccccc}
\hline
\textbf{\#Point} & \textbf{Total messages} & \textbf{Peers} & \textbf{CPU (\%)} & \textbf{How obtained} \\ \hline
1 & 1080.92 & 10  & 16.929 & Experimental \\ \hline
2 & 2474.27 & 20  & 18.651 & Experimental \\ \hline
3 & 3744.87  & 30  & 18.694 & Experimental \\ \hline
4 & 4747.98 & 40  & 20.645 & Experimental \\ \hline
5 & 6005.29 & 50  & 22.405 & Experimental \\ \hline
6 & 7568.48 & 60  & 22.924 & Experimental \\ \hline
7 & 8470.71 & 70  & 23.825 & Experimental \\ \hline
8 & 17527            & 142 & 32.620 & Regression   \\ \hline
9 & 24652            & 200 & 39.407 & Regression   \\ \hline
\end{tabular}
\end{table}

\section{Discussion and Future Work}
\label{sec:discussion}

%This paper focuses on improving the efficiency of intra-ledger communication for BFT blockchains, with XRP Ledger as a concrete use-case. 

Squelching is a solution designed by XRPL, and most parameters are originally fixed. However, we expect to explore in future some parameters such as \textit{timeout} or the \textit{node\_degree}. The number of peers a node keeps "alive" (does not squelch) at any given moment is an important aspect: a sparse network is more efficient w.r.t. the number of messages but less resilient, while a dense network will generate more duplicates, lower message latency, and will be more resilient~\cite{barabasi2016network}. Lowering too much the \textit{node\_degree} leads to security/resilience risks; too many active peers are "wasteful".

Even if our work is not focused on energy, the following can be considered: based on~\cite{Roma2020Energy}, saving 20\% CPU roughly translates to 5\% power saving. Noting that also memory and NIC would likely use less power when processing less data, we can expect savings better than 5\%. Besides, network traffic can make up a large portion of a Stellar node’s total consumption~\cite{9059429}. As Stellar shares design commonalities with XRP, there is a distinct possibility that network traffic takes a significant share of power in XRPL.

While Squelching is one solution to improve message dissemination, other alternatives for XRPL are discussed below.

\begin{figure}[ht!]
   			\begin{minipage}{0.44\textwidth}
   			\centering
       		\includegraphics[width=1\textwidth]{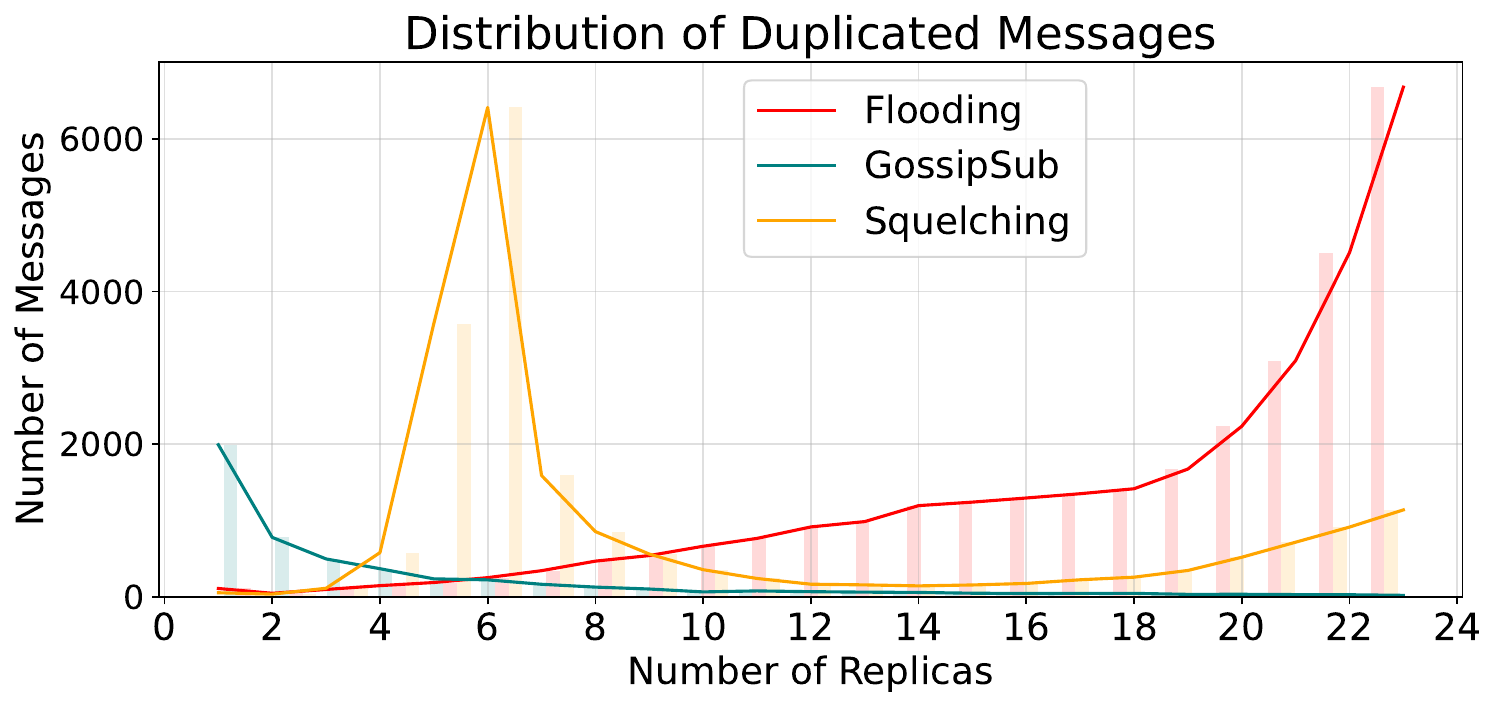}
			\caption{Frequency of duplicated messages on XRPL with Flooding vs Squelching vs Gossipsub~\cite{gossipsub-Flaviene}.}
            \label{fig: gossip_results}
        	\end{minipage}
		\end{figure}

\textit{(A) Gossipsub}~\cite{gossipsub-paper} is a publish-subscribe~\cite{pubSub} protocol enabling efficient and scalable message dissemination in P2P networks. Gossipsub nodes engage in a gossip-type protocol by selecting a sub-set of peers to share messages with, which then propagate them again recursively. There can be different "topics of discussion" between the nodes, for which different sets of peers are used at a given moment. The peer selection algorithm is complex and makes use of an extensive set of parameters. Gossipsub was originally proposed for Ethereum and Filecoin, and we previously proposed an adaptation for XRPL named FlexiPipe~\cite{gossipsub-Flaviene}. To facilitate the discussion, we partially reproduce a relevant result from~\cite{gossipsub-Flaviene} in Figure~\ref{fig: gossip_results}, where this solution was compared with \textit{squelching} and \textit{flooding}. It shows that Gossipsub is able to properly disseminate the validations with less duplication. However, this evaluation does not consider the overhead of messages generated to maintain the Gossipsub overlay.

Also, compared to Squelching, Gossipsub relies on many parameters for fine-tuning, and most of all, it assumes the deployment of a distinct overlay next to the native XRPL overlay. Actually, the main advantage of \textit{squelching} is to directly work over the XRPL overlay itself. The deployment is thus simplified and does not introduce possible external threats or failures from external overlays. 

\textit{(B) XRP-NDN.} Instead of sending data packets to specific locations (\textit{e.g.} IP addresses), Named Data Networking (NDN)~\cite{Brief_Intro_NDN,ndn-site,ndn} allows users to retrieve content by expressing what they want, similar to requesting a book by title: a user sends an interest packet with the data name. The forwarding is based on name prefixes instead of IP prefixes, and the request reaches one or multiple content providers or in-network caches with the expected content. Data is sent on the backward path and network nodes cache it distributively. 

In~\cite{XRP-NDN-overlay} we showed how the blockchain consensus messaging can be ported to use NDN for message propagation by proposing multiple mapping models for the forwarding of messages, and investigated the advantages and disadvantages of each model according to the specifics of XRPL. %Even if straightforward comparisons with squelching would not be fair due to the different nature of networks (\textit{e.g.} caching mechanisms which imply a distributed overhead in the network), performances reported in~\cite{XRP-NDN-overlay} are good for an NDN-based solution. 
However like Gossipsub, NDN implies an additional overlay network. %Besides, the size of the NDN interest message payload is limited, making the solution not viable for PoW-type blockchains.

Because in NDN data is named, another exploitable advantage is having a single \textit{validation} message per validator to enter each XRPL application. This can be done by filtering messages at NDN overlay level such that they do not reach the XRPL application, which can even use different hardware to take more advantage from filtering. The solution leaves open the possibility to use a structured overlay, or alternatively, an unstructured mesh mirroring the XRPL mesh. Naturally, a structured overlay could diminish the number of messages on the NDN overlay at the expense of security and robustness, while an overlay mirroring XRPL would exhibit the opposite.

While other solutions like dissemination trees may yield better results, they exhibit several drawbacks compared to flooding-like solutions as squelching or gossipsub: i) maintaining a tree structure is complicated and costly; ii) latencies increase as paths are longer; iii) topologies that change continuously force tree rebuilds which can impact latency and message delivery (service availability), and iv) resilience to attacks and node churning.

\begin{comment}
\subsection{Future work}

As highlighted above, solutions based on more structured overlay networks like Gossipsub or with complete changes in the message forwarding like NDN could offer large benefits to the BFT-based blockchains. The main issue is the creation of a side overlay network, with its own potential problems regarding failures, threats, or performance degradation. This is why the DLT community and businesses are quite reluctant to this kind of design choices. Rather than multiplying different overlay networks for a single application, it would be beneficial to integrate these protocols or communication architectures within the DLT protocols in an intrinsic manner. Obviously, this would require an in-depth security analysis because this might directly impact the core functionalities of a blockchain such as the consensus mechanism and its intrinsic security properties. 
\end{comment}

\section{Related Work}
\label{sec:relwork}

%others:

%-Squelching (Ripple)
% -Erlay: Efficient Transaction Relay for Bitcoin
%Reduce bw by 84% but increases latency for Tx dissemination by 2.6s (3.15-> 5.75s) 
%Perigee: Efficient Peer-to-Peer Network Design for Blockchains, Focuses on block propagation delay not flooding/bandwidth issue 
%GossipSub
%Epidemic Broadcast Trees https://core.ac.uk/download/pdf/32330596.pdf
%SplitStream
%Systematization of Knowledge:  Cross Chain Communication 
%https://docs.google.com/presentation/d/1WShriSfkq_V0SiLX9qVhNOji9CPcuV0R4ghzrxaGEpY/edit#slide=id.gbdb1ef01d0_0_624

%Hopefully - DONE - I gave more context. Just wanted to show that previous work addressed other topics, but id NOT address networking for XRPL and in general for consensus-validation based blockchain networks (XRPL is part of this family).
%\JF{Below there are four references about analysis of the consensus protocol introduced without really describing their outcomes. It is needed otherwise why citing them?}

%A literature review showed that previous work related to XRPL focused on topics different from ours, with little-to-none on-topic previous work: the XRPL consensus protocol was described in 2014~\cite{RPCA-old} with a main objective to challenge its security and robustness. 

XRP LCP was described in 2014~\cite{RPCA-old} followed by a rich literature investigating its robustness and security. It was analyzed in \cite{DBLP:journals/corr/abs-1802-07242,XRPL-protocol-improvement-small-world,consensus-algorithms} and investigated empirically in~\cite{xrpl-dfa}. This contrasts with our work focused on the message dissemination mechanism. In 2020, relatively simple cases were identified where consensus may violate safety and/or liveness~\cite{security_analysis_xrpl}, and it was argued that XRPL needs a very close \textit{synchronization}, \textit{interconnection}, and \textit{fault-free operation} between validators. This is another argument for considering also the efficiency and resilience of communication - the focus of our research. %A man-in-the-middle attack was demonstrated in~\cite{XRPL-MITM-attack}. Besides, XRPL-related work also investigates the energy consumption of the validators~\cite{XRPL-energy-waterloo}, the crypto-asset network flows~\cite{XRPL-cryptosasset-flows}, a health assessment of XRPL credit network~\cite{XRPL-credit-risk}, a proposal for a blockchain benchmarking framework~\cite{XRPL-benchmarking}, or a topology analysis~\cite{topo-xrpL}. 

While communication lacked in-depth exploration on XRPL and on consensus-validation-based blockchains in general, an efficient transaction relay for BTC named \textit{Erlay} was proposed in~\cite{erlay}. %It relies on two fundamental system-building strategies: delay and batching.
%Rather than disseminating each TX across every network link, a node selectively broadcasts it to a subset of its peers. TXs are universally propagated among well-connected public nodes through at most 8 outbound connections. Nodes regularly compare their states with their peers to consolidate them and only forward relevant information. 
While \textit{Erlay} reduces bandwidth by 84\%, the latency of TX dissemination increases from 3.15 to 5.75s on average, which is unacceptable on XRPL. \textit{Perigee}~\cite{perigee} is an efficient P2P network design for PoW blockchains focusing on mitigating the block propagation delay, but not on the message flooding issue. Epidemic Broadcast Trees are proposed on a gossip-based overlay in~\cite{epidemic-broadcast-trees} while \textit{Splitstream}~\cite{splitstream} distributes the load of forwarding messages evenly between participants. Actually, \textit{Gossipsub}~\cite{gossipsub-paper}, previously introduced, draws upon the general concepts of epidemic broadcast and gossip protocols. 

While \textit{XRP-NDN} was also reviewed here, there are other proposals to use NDN in the context of blockchains, however focusing on PoW type: \textit{BoNDN}~\cite{BoNDN}, proposes TX dissemination for BTC through a push model over NDN interests, and a subscribe-push model for block propagation. Another design for propagating TXs and blocks over NDN was proposed in~\cite{NDN-Ethereum} for ETH, while~\cite{BCXP} sends blocks over a multi-layer design based on NDN to achieve 74\% of BlockNDN's overhead~\cite{BlockNDN}. Being content-oriented, NDN is suited for data synchronization and different protocols were proposed: \textit{Vectorsync} \cite{vectorsync}, \textit{Chronosync} \cite{chronosync}, \textit{Psync} \cite{pSync}. %These are good candidates to be leveraged in a blockchain context where states across multiple nodes must be consolidated, and \textit{BlockNDN} actually relies on \textit{Chronosync}. 
However, they were not meant originally for the byzantine blockchain environment~\cite{NDN-Ethereum}; moreover, here we want to minimize the number of messages, and the additional sync messages can add unwanted overhead.

\section{Conclusion}

In this paper, we formulated the problem of \textit{Node}-wise scalability in the context of XRPL to highlight the rationale behind the Squelching approach. Our empirical approach was first focused on measuring and modeling the node overhead in regard to the number of messages disseminated in the XRPL P2P network. As they represent the vast majority of messages, squelching \textit{validation} and \textit{proposal} messages, as proposed in the XRPL node software, is beneficial. We evaluated its gain using a regression model to understand its impact on the production XRPL network, from measurements done in a controllable environment. An XRPL hub node can potentially benefit from a 29\% connectivity increase when using \textit{Squelching}.

At broad level, it is expected that with respect to message efficiency over flooding, \textit{squelching} can benefit other blockchains too. For instance, in~\cite{Lokhava2019FastAS}, the current broadcast of messages in STELLAR is inefficient and many scaling issues can be addressed by optimizing network traffic. By using Gossipsub, the combination of node-degree tweaking and epidemic broadcast improved ETH communication layer. 

%Compared to flooding, message propagation techniques reducing the amount of network traffic and computational resources required to disseminate information make a network more scalable, to both handle a larger number of transactions and accommodate more nodes.
Because Squelching might impact the robustness and security of the consensus mechanism itself, future work will thus consider a thorough security evaluation.

\label{sec:conclusion}
%\input{outline}
%\begin{comment}

\section*{Acknowledgment}
%
%%This work was funded in part by the
This research was funded in whole, or in part, by the Luxembourg National Research Fund (FNR), grant references 
 PRIDE15/10621687/SPsquared and INTER/14783140/GLADIS. For the purpose of open access, and in fulfilment of the obligations arising from the grant agreement, the author has applied a Creative Commons Attribution 4.0 International (CC BY 4.0) license to any Author Accepted Manuscript version arising from this submission. In addition, we thankfully acknowledge the support from the RIPPLE University Blockchain Research Initiative 
(UBRI) for our research.  

%\end{comment}

%The preferred spelling of the word ``acknowledgment'' in America is without an ``e'' after the ``g''. Avoid the stilted expression ``one of us (R. B. G.) thanks $\ldots$''. Instead, try ``R. B. G. thanks$\ldots$''. Put sponsor acknowledgments in the unnumbered footnote on the first page.

\bibliographystyle{style/IEEEtran}
\bibliography{bib/ebpf}

\end{document}